\documentclass[journal=jacsat,manuscript=article]{achemso}

\usepackage[version=3]{mhchem} % Formula subscripts using \ce{}
\usepackage{xcolor}
\usepackage{soul}
\usepackage{microtype} % fixing underfull and overfull problem
\definecolor{myblue}{HTML}{2d509c}
\usepackage[colorlinks=true, linkcolor=myblue, citecolor=black, urlcolor=myblue]{hyperref}
\usepackage{float}
\usepackage{array}
\usepackage{subcaption}
\usepackage{multirow}
\usepackage{makecell}
\usepackage{minted}
\usepackage{indentfirst}
\usepackage{longtable}
\usepackage{orcidlink}

\author{Aritra Roy\orcidlink{0000-0003-0243-9124}}
\affiliation
{Energy, Materials and Environment Research Centre, London South Bank University, London SE1 0AA, UK.}
\alsoaffiliation{School of Engineering and Design, London South Bank University, London SE1 0AA, UK.}
\email{pgr.aritra.roy@lsbu.ac.uk}
\author{Enrico Grisan\orcidlink{0000-0002-7365-5652}}
\affiliation{Bioscience and Bioengineering Research Centre, London South Bank University, London SE1 0AA, UK.}
\author{John Buckeridge\orcidlink{0000-0002-2537-5082}}
\affiliation
{Energy, Materials and Environment Research Centre, London South Bank University, London SE1 0AA, UK.}
\alsoaffiliation{School of Engineering and Design, London South Bank University, London SE1 0AA, UK.}
\email{j.buckeridge@lsbu.ac.uk}
\author{Chiara Gattinoni\orcidlink{0000-0002-3376-6374}}
\affiliation{Department of Physics, Kings College London, London WC2R 2LS, UK.}
\email{chiara.gattinoni@kcl.ac.uk}

\title[An \textsf{achemso} demo]
  {ComProScanner: A multi-agent based framework for composition-property structured data extraction from scientific literature }

\abbreviations{IR,NMR,UV}

\begin{document}

% \begin{tocentry}

% Some journals require a graphical entry for the Table of Contents.
% This should be laid out ``print ready so that the sizing of the

% Inside the \texttt{tocentry} environment, the font used is Helvetica
% 8\,pt, as required by \emph{Journal of the American Chemical

% The surrounding frame is 9\,cm by 3.5\,cm, which is the maximum
% permitted for  \emph{Journal of the American Chemical Society}
% graphical table of content entries. The box will not resize if the
% content is too big: instead it will overflow the edge of the box.

% This box and the associated title will always be printed on a
% separate page at the end of the document.

% \end{tocentry}

\begin{abstract}
Since the advent of various pre-trained large language models, extracting structured knowledge from scientific text has experienced a revolutionary change compared with traditional machine learning or natural language processing techniques. Despite these advances, accessible automated tools that allow users to construct, validate, and visualise datasets from scientific literature extraction remain scarce. We therefore developed \textsf{ComProScanner}, an autonomous multi-agent platform that facilitates the extraction, validation, classification, and visualisation of machine-readable chemical compositions and properties, integrated with synthesis data from journal articles for comprehensive database creation. We evaluated our framework using 100 journal articles against 10 different LLMs, including both open-source and proprietary models, to extract highly complex compositions associated with ceramic piezoelectric materials and corresponding piezoelectric strain coefficients (\textit{d$_{33}$}), motivated by the lack of a large dataset for such materials. DeepSeek-V3-0324 outperformed all models with a significant overall accuracy of 0.82. This framework provides a simple, user-friendly, readily-usable package for extracting highly complex experimental data buried in the literature to build machine learning or deep learning datasets.
\end{abstract}

\section{Introduction}\label{sec:introduction}
Contemporary data-driven materials design heavily relies on high-fidelity datasets in machine-readable formats, as the effectiveness of machine learning (ML) and deep learning (DL) methodologies hinges on structured and computationally accessible data containing, at minimum, material compositions and their corresponding physical properties. Over the past decade and a half, the establishment of computational databases of high-throughput screened materials based on Density Functional Theory (DFT) calculations, such as the Materials Project (MP)\cite{jain2013commentary}, JARVIS-DFT\cite{choudhary2020joint}, and Open Quantum Materials Database (OQMD)\cite{saal2013materials}, together with experimental datasets like the Cambridge Crystallographic Data Centre (CCDC)\cite{allen2002cambridge} or High Throughput Experimental Materials (HTEM) database\cite{zakutayev2018open}, has shifted research emphasis toward data-driven materials design. Nevertheless, the preponderance of experimental scientific knowledge regarding solid-state materials, analogous to other domains, remains embedded within millions of scientific journal articles. Extracting this wealth of information into the structured, machine-readable formats required for computational analysis presents a significant challenge that necessitates automated approaches.

Natural language processing (NLP) algorithms have demonstrated remarkable advances in materials science applications, from building toolkits and techniques for automated extraction of chemical information from the scientific literature, such as ChemDataExtractor\cite{swain2016chemdataextractor}, \cite{jessop2011oscar4}, ChemicalTagger\cite{hawizy2011chemicaltagger}, BatteryBERT \cite{huang2022batterybert}, and others. These tools and techniques have been implemented to systematically structure the vast corpus of textual knowledge in the field\cite{kim2017materials, kononova2019text, huang2020database, sierepeklis2022thermoelectric, dong2022auto, cruse2022text, beard2022perovskite, trewartha2022quantifying, huang2022batterybert, foppiano2023automatic} leveraging various techniques, including regular expressions\cite{python_regex}, BiLSTM recurrent neural networks\cite{sharfuddin2018deep}, and smaller transformer-based language models like BERT\cite{devlin2019bertpretrainingdeepbidirectional}. These approaches have successfully facilitated the extraction of entity information from diverse sources, including battery materials literature\cite{huang2020database, huang2022batterybert} and chemical synthesis parameters documented in methodology sections of scientific papers\cite{kononova2019text}. Entity extraction, and in particular named entity recognition (NER), has dominated these research efforts. Researchers have applied domain-specific labels such as ``material'' or ``property'' to specific textual elements, but require an additional post-processing step to construct the relations between these entities, relations that prove essential for training effective machine learning or deep learning models. To exemplify, discrete entities such as ``\ce{Cu2O}'' or ``2.1 eV'' were targeted rather than establishing the relational connections between them (for example, ``2.1 eV'' represents the measurement of the band gap for ``\ce{Cu2O}''), i.e., they do not implement relation extraction (RE) techniques. 

In the early 2020s, several end-to-end methods were developed that use a single machine learning model integrating both named entity recognition and relation extraction (NERRE)\cite{giorgi2022sequence, cabot2021rebel, townsend2021doc2dictinformationextractiontext}. These methodologies demonstrate efficacy in relation extraction tasks; however, they remain fundamentally limited to \textit{n}-ary relation extraction frameworks that are complex in architectural structure and struggle to extract all information if the interconnection between various entities are too high. Following the widespread adoption of various large language models (LLMs), researchers have employed them successfully to extract information from journal articles, replacing traditional sequence-to-sequence approaches with more sophisticated NERRE methods. Approaches ranging from pre-training\cite{mishra2024llamat} and fine-tuning LLMs\cite{dagdelen2024structured, foppiano2024mining, polak2024flexible, ye2024constructionapplicationmaterialsknowledge, jablonka202314, zimmermann2025reflections2024largelanguage} to prompt-engineering\cite{polak2023extracting, alampara2025probinglimitationsmultimodallanguage, prasad2024developmentautomatedknowledgemaps, foppiano2024mining, gupta2024data, jablonka202314, zimmermann2025reflections2024largelanguage}, zero-shot\cite{foppiano2024mining, polak2024flexible, ekuma2024dynamicincontextlearningconversational, jablonka202314, zimmermann2025reflections2024largelanguage} and few-shot prompting\cite{foppiano2024mining, ekuma2024dynamicincontextlearningconversational, gupta2024data, zimmermann2025reflections2024largelanguage}, as well as Retrieval-Augmented Generation
(RAG) methods\cite{lala2023paperqa, maharana2025retrieval} have enhanced NERRE-level text extraction from materials science literature. Concurrently, LLM-powered agents have been utilised for various chemistry and material science tasks, including extracting relevant information from journal articles\cite{m2024augmenting, ansari2024agent, zhang2024honeycombflexiblellmbasedagent, skarlinski2024languageagentsachievesuperhuman, feng2025agentic}, predicting new molecules or materials or their properties\cite{zimmermann2025reflections2024largelanguage, m2024augmenting}, automating data handling\cite{zimmermann2025reflections2024largelanguage, zhang2024honeycombflexiblellmbasedagent, chiang2024llamplargelanguagemodel, boiko2023autonomous, feng2025agentic}, enhancing reasoning and computational capabilities of LLMs\cite{m2024augmenting, zhang2024honeycombflexiblellmbasedagent, skarlinski2024languageagentsachievesuperhuman, chiang2024llamplargelanguagemodel, boiko2023autonomous}, proposing novel hypotheses\cite{zimmermann2025reflections2024largelanguage}, and even semi-automating experiments\cite{boiko2023autonomous} by integrating expert tools. Several notable implementations have emerged in this domain, such as Eunomia by Ansari \textit{et al.}\cite{ansari2024agent}, an AI agent chemist for developing materials datasets by accessing computational databases and research papers, and, very recently the multi-agent system nanoMINER\cite{odobesku2025agent}, which combines LLMs and multimodal analysis to extract information, though it is specifically limited to nanomaterials. However, both Eunomia and nanoMINER lack the capability to integrate Text and Data Mining (TDM) API keys\footnote{TDM agreements differ from standard academic subscriptions granted to institutional libraries, as they specifically govern the scraping and downloading of large volumes of content, which could potentially impact the operational performance of publishers' servers.} through the package, requiring users to provide the articles in PDF format by manually downloading them, which represents a labour-intensive and time-consuming process when dealing with large-scale datasets. Additionally,enumerating all explicit chemical formulas from variable compositions (e.g., Pb$_{1-x}$K$_x$Nb$_2$O$_6$ where x=0.1, 0.2 etc.) into distinct compounds remains beyond the scope of these agentic systems. Recently, Wilhelmi \textit{et al.} published a comprehensive tutorial on using LLMs to extract chemical data as structured output via various methods, including prompting, RAG and agentic systems\cite{schilling2025text}. Nevertheless, an easily configurable automated workflow that enables end users to build, evaluate, and visualise datasets through information extraction from journal articles has been lacking.

\renewcommand{\thefootnote}{\roman{footnote}}
In this work, we present an autonomous multi-agent agile framework, \textsf{ComProScanner}, for end users to extract, evaluate, categorise and visualise machine-readable structured chemical compositions and properties, combined with synthesis information from journal articles to create extensive databases. When a research article contains chemical composition along with the enquired property value, the framework extracts structured JSON data\cite{json_website} containing both agent-extracted relevant information and journal article metadata obtained via APIs. The agent-extracted relevant information comprises the chemical composition of the material and the property value as key-value pairs, property unit, material family, synthesis method, precursors used, brief synthesis steps, and characterisation techniques employed. Our system combines LLM agents with powerful tools, including RAG and a custom deep learning model for extracting chemical compositions and properties only when property values are available in articles. The workflow supports Elsevier, Springer Nature, IOP Publishing, and Wiley articles via publishers' TDM APIs or PDFs from local folders. \textsf{ComProScanner} enhances text-mining accuracy by providing flexible contextual parameters to agents while maintaining cost-effectiveness through preliminary article filtering via keyword matching. The system supports multiple configurable LLMs for both extraction agents and RAG implementations. \textsf{ComProScanner} can be implemented with fewer than 20 lines of Python code to extract pre-defined structured data, provided that users have access to the TDM APIs of the publishers. We evaluated the extraction performance of ten LLMs using 100 articles containing piezoelectric coefficient \textit{d$_{33}$} values, achieving overall accuracy exceeding 80\% across various models. Detailed evaluation methods and metrics are presented in the Results and Discussion sections. Additionally, we conducted cost versus accuracy comparisons to help users identify economically suitable LLMs to build datasets without performing their own evaluations.

% \begin{figure}[!t]
%     \centering
%     \includegraphics[width=1\linewidth]{images/example_json.png}
%     \caption{Example of structured JSON output for a scientific article using the  \textsf{ComProScanner} package for piezoelectric property data.}
%     \label{fig:example_json}
% \end{figure}

\section{Methods}\label{sec:methods}
\textsf{ComProScanner} is a highly configurable multiagent-based Python package developed using CrewAI\cite{moura2023crewai}, a production-grade framework for orchestrating AI agent workflows,  supplemented with custom Python scripts. Custom scripts have been strategically implemented throughout the system to enhance cost-effectiveness and accessibility for researchers engaged in data-driven materials discovery. 

\begin{figure}
    \centering
    \includegraphics[width=1\linewidth]{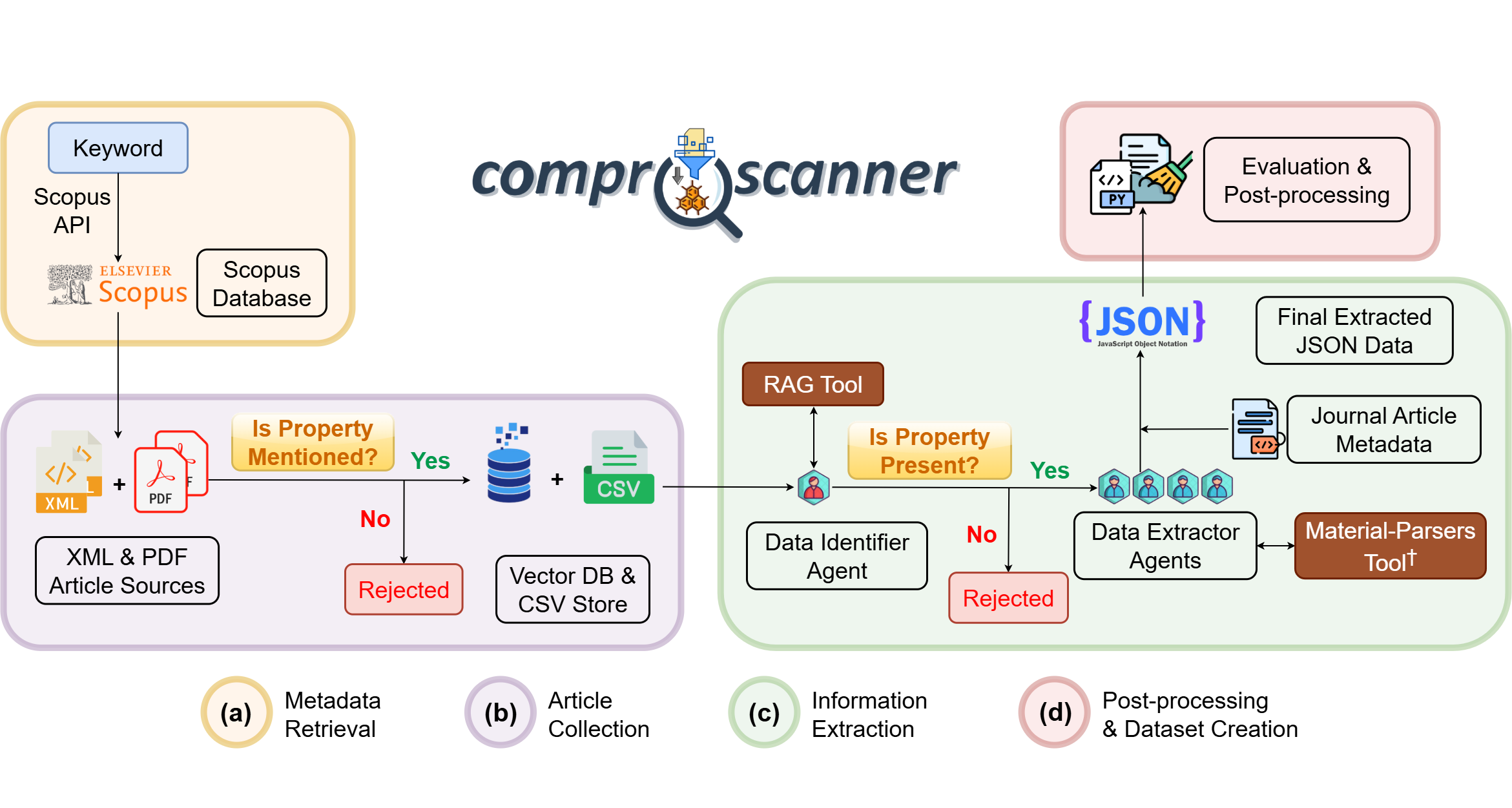}
    \caption{Overall workflow diagram of \textsf{ComProScanner} framework, separated in four distinct operational phases, distinguished with four different colour regions: (a) metadata retrieval (yellow), (b) article collection (purple), (c) information extraction (green), and (d) evaluation, post-processing, and dataset creation (brown).}
    \label{fig:overall_workflow}
\end{figure}

\textsf{ComProScanner}'s workflow architecture comprises four distinct operational phases: (a) metadata retrieval, (b) article collection, (c) information extraction, and (d) evaluation, post-processing, and dataset creation (see \autoref{fig:overall_workflow}). We describe each phase in turn below.

\subsection{Metadata Retrieval}\label{subsec:metadata-retrieval}
In this phase, \textsf{ComProScanner} finds metadata for relevant articles associated with the enquired property, including DOI, publication name, ISSN, Scopus ID, article title, article type, and publisher name. This section of the program implements property-related article metadata retrieval scripts that function as a Python wrapper for the Scopus Search API\cite{scopus_search_api}. The wrapper enables users to specify primary keyword(s) for relevant metadata search while providing the flexibility to incorporate additional keywords in combination with the primary terms. In alignment with the objectives of the \textsf{ComProScanner} package, this module filters the document formats to include only \textit{Articles} and \textit{Letters}, thereby eliminating other document types such as \textit{Reviews} or \textit{Conference Papers} that could potentially introduce duplicate compositions or properties into the dataset.

\subsection{Article Collection}\label{subsec:article-collection}
This section of \textsf{ComProScanner} accesses full-text articles through publisher-provided TDM APIs. The system currently supports automated extraction of articles from four major publishers via their TDM API and manually downloaded PDF articles from all publishers (for further details, see section S1 of the Electronic Supplementary Information (ESI)). The system implements preliminary keyword-based filtration for the entire text of the article through Python regular expressions (Python RegEx)\cite{python_regex} to identify relevant articles mentioning the property, thereby optimising data management by avoiding text extraction from irrelevant sources that would unnecessarily inflate the database size. Articles in which the required property is mentioned are organised and stored in CSV format with dedicated columns corresponding to specific article sections (abstract, introduction, experimental methods, computational methods, results and discussion, and conclusion), with optional MySQL database\cite{MySQL2025} integration. Vector databases are generated, along with CSV files, using the open-source ChromaDB\cite{Chroma2025} package when any of the specified relevant keywords are detected within an article, facilitating future RAG queries.

\begin{figure}
    \centering
    \includegraphics[width=0.85\linewidth]{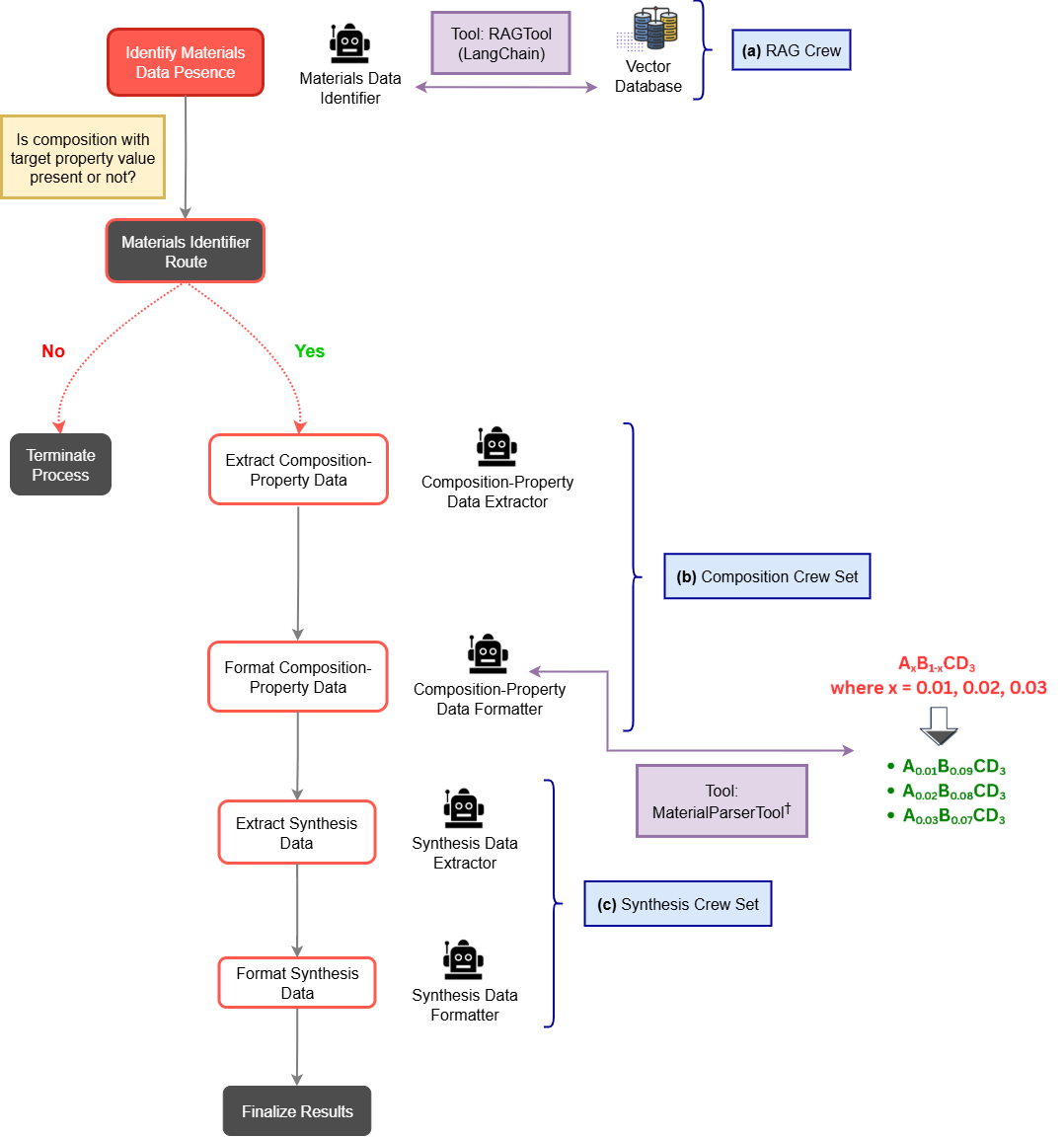}
    \caption{Comprehensive workflow diagram of the CrewAI-based extraction system in ComProScanner, comprising five specialised agents. The process begins with a property identifier agent ((a) RAG Crew) that leverages Retrieval-Augmented Generation (RAG) technology to filter relevant articles. The remaining four agents are strategically organised into two parallel functional subgroups: one dedicated to composition data extraction ((b) Composition Crew Set) and the other focused on synthesis information collection ((c) Synthesis Crew Set). Each subgroup implements a sequential two-agent architecture—the first agent extracts raw data while the second performs formatting and standardisation. The workflow integrates two essential tools: RAGTool for discriminating between mere property mentions and actual quantitative property values, and MaterialParserTool for accurate processing of complex chemical formulations.}
    \label{fig:flow_diagram}
\end{figure}

\subsection{Information Extraction}\label{subsec:information-extraction}
The information extraction phase represented in detail in \autoref{fig:flow_diagram}, incorporates five specialised AI agents (\autoref{fig:flow_diagram}), beginning with a property identifier (the Materials Data identifier) that utilises RAG technology under the RAG Crew. This initial filtering significantly reduces API costs (or computational resource usage for locally hosted LLMs) by eliminating articles that merely mention the required property without containing actual property values. The four remaining agents are organised into two functional subgroups: one dedicated to extracting composition-related data (the Composition Crew Set) and another focused on collecting synthesis information (the Synthesis Crew Set). Each subgroup employs two sequentially ordered agents; the first extracts raw data, while the second formats it. By default, the package uses the provided keyword with some pre-set rules and instruction for the agents. However, \textit{Notes} can be appended to both agents and tasks across all five agent components to provide supplementary instructions to the agents as additional context. For complex compositions with multiple fractions denoted as variables e.g., \textsf{Na$_{(1-x)}$Li$_x$TiO$_3$} where \textsf{x=0.1, 0.3, 0.4}, the system employs material-parsers, a deep learning model developed by Foppiano \textit{et al.}\cite{foppiano2023automatic}, as an agent tool that resolves the example into three distinct compositions: \textsf{Na(0.9)Li(0.1)TiO3}, \textsf{Na(0.7)Li(0.3)TiO3}, and \textsf{Na(0.6)Li(0.4)TiO3}. All extracted data are compiled into a unified JSON format, which is subsequently integrated with the corresponding article metadata\footnote{This new article metadata is collected for each specific article containing agent-extracted information, differing from the previously collected metadata that contained limited information for all related articles associated with the property keyword used for metadata collection. This new comprehensive metadata includes a wide range of information: DOI, article title, journal name, year of publication, open access information, author list with their institutional details, and article keywords. These data are obtained either via Elsevier's ScienceDirect Article Metadata API\cite{sciencedirect_metadata_api} (optional) or the Open Access Button's free metadata API\cite{openaccess_metadata_api} developed by OA.Works\cite{oaworks_website}}.

\subsection{Evaluation, Post-processing, and Dataset Creation}\label{subsec:evaluation-post-processing}

The total extracted JSON data comprise two main segments: (i) agent-extracted composition-property and synthesis data, (ii) journal article metadata. A detailed description of each type of agent-extracted data can be found in section S2 of the ESI along with an example of complete extracted JSON data (Figure S1). \textsf{ComProScanner} offers a built-in comprehensive evaluation framework, both agent-based and semantic-based, designed to assess and visualise the extraction performance of LLM agents with scientific rigour. The framework implements three distinct categories of evaluation metrics: (a) custom weight-based accuracy metrics, (b) conventional classification metrics, and (c) normalised classification metrics. The custom weight-based accuracy enables users to assign differential priority values to specific extracted parameters; for instance, users can allocate greater emphasis to composition-property key-value pairs (weight of 0.3) compared to synthesis steps (weight of 0.1) [see section S2 in the ESI for all possible extracted parameters and their weight-based emphasis], where the total weight for all extracted parameters is 1. If the weight for one parameter (e.g., composition-property) is set to 1, keeping all other parameters 0, the accuracy will be determined purely by the composition-property extraction performance. Standard classification metrics include Precision, Recall and F1-score, which are defined relative to the concepts of true positive (TP), false positive (FP), and false negative (FN). True positive can be defined as a correct value extracted by the agent, false positive when the extracted value does not match the ground truth (the actual text to be extracted), and false negative as a value that is expected but has not been extracted by the agent. Both Precision and Recall can be represented as:
\begin{equation}
    \text{Precision} = \frac{TP}{TP+ FP}\text{, }
    \text{Recall} = \frac{TP}{TP + FN}
    \label{eq:precision_recall}
\end{equation}
From Precision and Recall, another metric, F1, can be calculated using \autoref{eq:f1_eq},
\begin{equation}
    \text{F1} = \frac{\text{2}\times \text{\textit{Precision}}\times\text{\textit{Recall}}}{\text{\textit{Precision} + \textit{Recall}}}
    \label{eq:f1_eq}
\end{equation}
Beyond standard classification metrics calculated using the aggregate number of items across all evaluation articles, we have developed normalised classification metrics that consider each article as a single evaluative unit, wherein each extracted item within an article contributes a fractional importance to that article's overall evaluation score. These normalised evaluation metrics were specifically designed to ensure an equitable comparison between articles with significant disparities in the quantity of extractable information. The normalised metrics for all papers are calculated using the modified Precision, Recall and F1-score,
\begin{equation}
   \text{Normalised Precision} = \frac{\sum_{i=1}^{N} (\frac{TP_{i}}{TP_{i}+FP_i})n_i}{N}
\label{eq:normalised_precision}
\end{equation}
\begin{equation}
   \text{Normalised Recall} = \frac{\sum_{i=1}^{N} (\frac{TP_{i}}{TP_{i}+FN_i})n_i}{N}
\label{eq:normalised_recall}
\end{equation}
\begin{equation}
    \text{Normalised F1} = \frac{\text{2}\times \text{\textit{Normalised Precision}}\times\text{\textit{Normalised Recall}}}{\text{\textit{Normalised Precision} + \textit{Normalised Recall}}}
    \label{eq:normalised_f1}
\end{equation}
where, $TP_i, FP_i, FN_i$ = true positives, false positives, false negatives for paper i, $n_i$ = total number of items in paper \textit{i}, and $N$ = total number of papers.

Weight-based accuracy metrics, classification metrics and normalised classification metrics all provide the flexibility to use both semantic and agentic approaches for evaluation. The semantic similarity method is used to match ground truth and \textsf{ComProScanner}-extracted information for the semantic approach, whereas LLM agents are instructed to match the ground truth and \textsf{ComProScanner}-extracted information for the agentic approach. Although the evaluation accuracy is expected to be higher for the agentic approach, given that LLM agents will have better comparison ability than semantic comparison between two sentences, the agentic evaluation can take more time and require significantly large numbers of tokens if reasoning models are used for better performance.

\textsf{ComProScanner} provides extensive visualisation capabilities of the evaluation through a diverse array of graphical representations, including bar charts, radar plots, heat maps, histograms, and violin charts, all readily accessible within the framework. Additionally, the system offers pie charts and histogram plotting functionalities to facilitate the analysis of data distribution across composition families, precursors, and characterisation techniques.

\section{Results}\label{sec:results-discussion}
Although the Materials Project\cite{jain2013commentary} contains the largest database of piezoelectric materials\cite{de2015database}, approximately 700 materials therein have non-zero \textit{d$_{33}$} coefficients, which quantify the electric field generated when a piezoelectric material is subjected to applied strain. More critically, fewer than 50 materials are present in the database with \textit{d$_{33}$} values exceeding 10 pC/N, where the highest value reaches up to 738.47 pC/N. However, tens of thousands of previous works, predominantly experimental, have been already conducted to identify materials with larger \textit{d$_{33}$} coefficients through doping or other methods, yet these findings remain in the literature in unstructured, non-machine readable formats. Thus, considering the challenge of extracting composition-property relationships from unstructured literature data as one of the most significant tests for assessing \textsf{ComProScanner}'s ability, along with synthesis data, we evaluated ceramic piezoelectric materials and their corresponding piezoelectric \textit{d$_{33}$} coefficient values, across ten different LLMs.

Metadata of articles related to piezoelectric materials were collected based on \textit{piezoelectric}, \textit{piezoelectricity}, \textit{pyroelectric}, \textit{pyroelectricity}, \textit{ferroelectric}, and \textit{ferroelectricity} as the main base keywords. After collecting metadata with only base queries, combinations of base queries and 18 additional keywords such as, \textit{advancements}, \textit{applications}, \textit{ceramics}, \textit{characterization}, \textit{composites}, \textit{crystals}, etc., were used to collect a larger set of metadata that could contain potential piezoelectric materials along with their corresponding \textit{d$_{33}$} coefficient values. The complete list of the additional keywords can be found in the supporting information\cite{roy2025comproscanner} (see the test\_example.py script). Although metadata were collected for all articles published between 1st of January 2019 and 17th of March 2025, only Elsevier papers were considered for the evaluation process, where only 3,916 papers mentioned \textit{d$_{33}$}, accounting for potential differences in formatting. 

Subsequently, 100 test DOIs were selected, based on the presence of the composition-property data using the RAG agent, whilst randomising the metadata order. For RAG and other NLP tasks, text embedding plays a crucial part in ensuring the efficiency of the models. The PhysBERT\cite{hellert2024physberttextembeddingmodel} model has demonstrated superior accuracy compared to various sentence transformer and BERT models in identifying various physics and materials science specific vocabulary. However, to  ensure that PhysBERT would perform better than the leading sentence transformer model, all-mpnet-base-v2\cite{sentencetransformers2021mpnet}, in our specific domain, the \textit{thellert/physbert\_cased} model from Hugging Face was evaluated against sentence-transformer's all-mpnet-base-v2 model using 12 domain-specific synonyms based on abbreviated forms or chemical formulae and their corresponding full names or trivial names, which are summarised in Table S1 in the ESI. The PhysBERT model outperformed all-mpnet-base-v2 in all cases, with remarkable performance differences ranging from highly significant improvements for terms such as DOS (density of states) with a cosine similarity\footnote{Cosine similarity measures how similar two text embeddings (vectors) are by calculating the cosine of the angle between them. This provides a score between -1 (completely dissimilar) and 1 (identical), which is useful for tasks such as text or document clustering.} difference of 0.8338, to modest improvements for common terms such as PVC (polyvinyl chloride) with a difference of 0.0566. This satisfactory performance of PhysBERT encouraged us to adopt this model as the default embedding model for storing article text data in the ChromaDB vector database for use in the RAG tool illustrated in \autoref{fig:flow_diagram}. For fair evaluation across various models, the RAG environment was maintained consistently, as described in detail in section S3 of the ESI.

Finally, extraction agents were used to extract information for the piezoelectric materials only for the test DOIs. We selected LLMs to enable a comparison between open-source models (Google's Gemma-3-27B-Instruct\cite{gemmateam2025gemma3technicalreport}, DeepSeek's DeepSeek-V3-0324\cite{deepseekv3}, Meta's Llama-3.3-70B-Instruct, Llama-4-Maverick-17B-Instruct\cite{grattafiori2024llama3herdmodels, llama4}, and Alibaba's Qwen3-235-A22B\cite{qwen3technicalreport}, Qwen-2.5-72B-Instruct\cite{qwen2025qwen25technicalreport}) and proprietary models (Google's Gemini-2.0-Flash\cite{google2024gemini20introducing}, Gemini-2.5-Flash-Preview\cite{comanici2025gemini25pushingfrontier}, and OpenAI's GPT-4o-mini\cite{openai2024gpt4technicalreport}, GPT-4.1-nano)\cite{openai2025gpt41}) at similar price points, after analysing the cost-versus-accuracy ratio from the Chatbot Arena LLM Leaderboard\cite{chiang2024chatbot}, where models had Arena scores exceeding 1250 and output costs below \$1/1M tokens (for more details, see section S4 in the ESI). Additional instructions were passed to the agents for better extraction performance specific to piezoelectric materials and \textit{d$_{33}$} coefficients (see the test\_example.py script from supporting information\cite{roy2025comproscanner}).

\begin{figure}[h]
    \centering
    \includegraphics[width=1\linewidth]{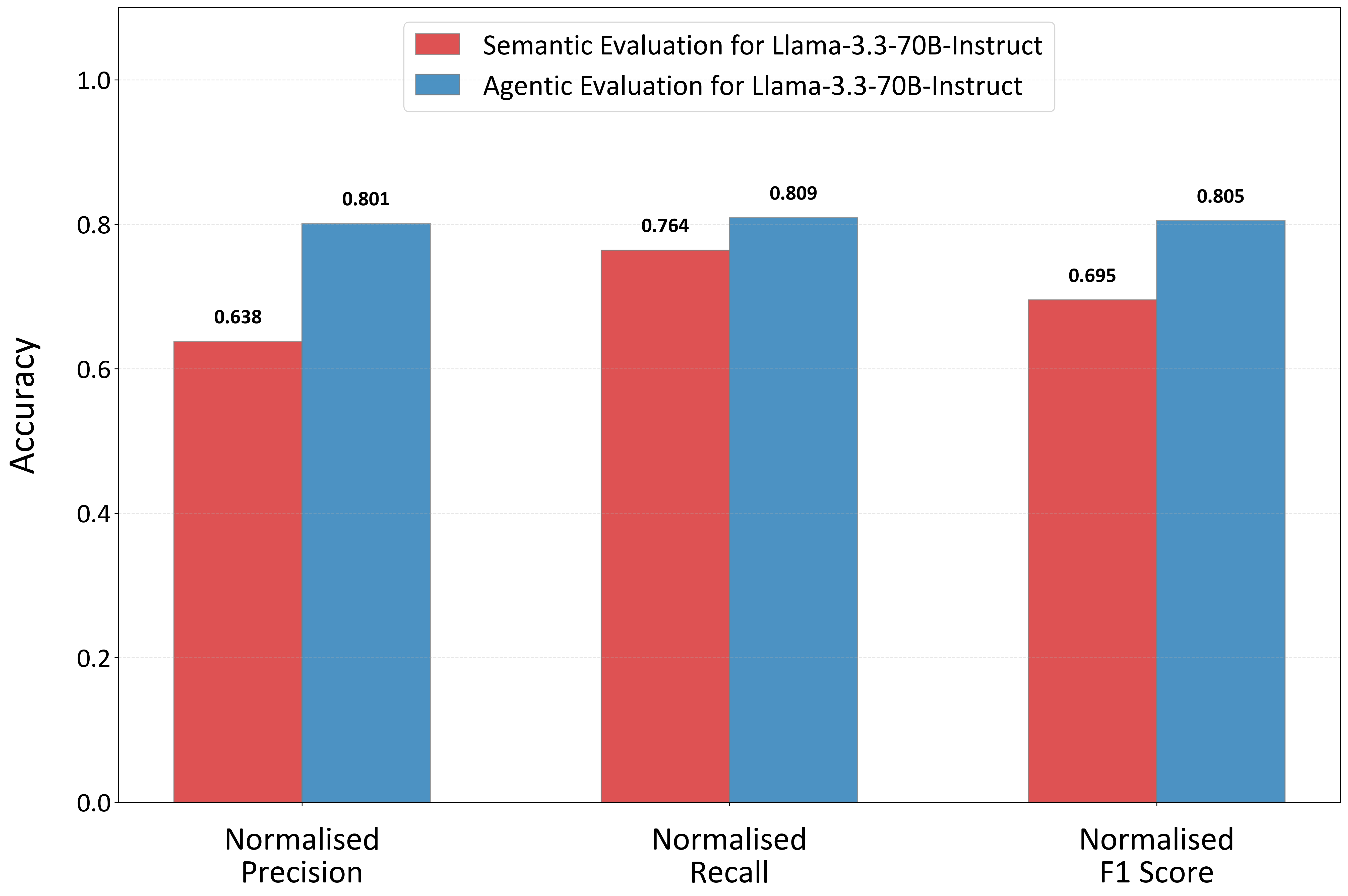}
    \caption{Normalised classification metrics (Precision, Recall, and F1-score) for model Llama-3.3-70B-Instruct (best performing model considering only normalised metrics) to showcase the performance of \textsf{ComProScanner}'s performance capability for the considered models using: semantic evaluation through the PhysBERT model (red bars), and agentic evaluation through the Gemini-2.5-Pro reasoning model (blue bars).}
    \label{fig:single_semantic_vs_agentic_comparison}
\end{figure}

The normalised classification metrics (Precision, Recall, and F1-score) of different models, as described in the \nameref{subsec:evaluation-post-processing} sub-section above, for both semantic and agentic approaches, are represented as grouped bar charts for the model Llama-3.3-70B-Instruct (the best-performing model when considering only normalised metrics) in \autoref{fig:single_semantic_vs_agentic_comparison}. Semantic and agentic comparisons based on normalised metrics for all other models can be found in section S5 of the ESI. We used PhysBERT model for semantic evaluation, the Gemini-2.5-Pro reasoning model\cite{comanici2025gemini25pushingfrontier} has been employed for agentic evaluation. Although normalised classification metrics for both semantic and agentic evaluation show similar trends, the agentic evaluation demonstrates superior performance accuracy compared to semantic evaluation, which is understandable given that reasoning models such as Gemini-2.5-Pro possess greater capability to compare sentence structures with equivalent meanings. Given the superior accuracy demonstrated by agentic evaluation, we focus on these results to identify the best-performing models for practical implementation. As mentioned earlier, Llama-3.3-70B-Instruct outperforms all other models in normalised classification metrics with a Precision value of 0.80, Recall value of 0.81, and F1-score of 0.80 (\autoref{fig:single_semantic_vs_agentic_comparison}). 

\begin{figure}[t!]
    \centering
    \includegraphics[width=1\linewidth]{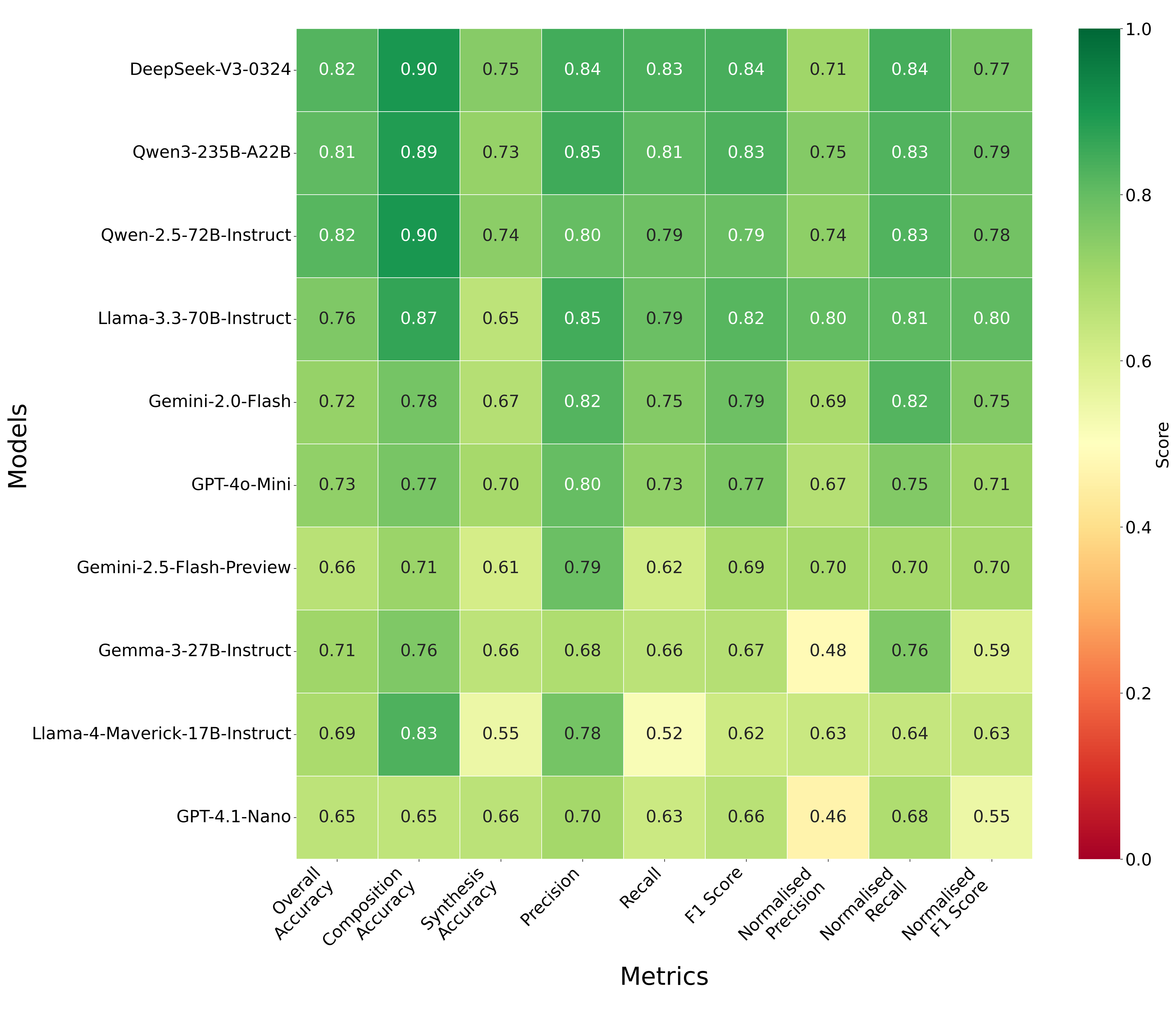}
    \caption{Confusion matrix from agentic evaluation, showcasing all 9 evaluation parameters, such as weight-based overall accuracy (average of composition accuracy and synthesis accuracy), weight-based composition accuracy and weight-based synthesis accuracy, classification metrics (Precision, Recall, and F1-score), and normalised classification metrics (normalised Precision, normalised Recall, and normalised F1-score), across 10 different LLMs used in this study.}
    \label{fig:confusion_matrix}
\end{figure}

The confusion matrix (\autoref{fig:confusion_matrix}) reveals distinct performance patterns across the evaluated models for piezoelectric materials extraction taking into account all performance metrics. DeepSeek-V3-0324 emerged as the top-performing model for data extraction, demonstrating consistently high scores across all metrics, with particularly strong performance in composition accuracy (0.90), Precision (0.84), Recall (0.83), and F1-score (0.84). This model showed balanced performance with an overall accuracy of 0.82 and robust synthesis accuracy of 0.75. The Qwen model family demonstrated competitive performance, with both Qwen3-235B-A22B and Qwen-2.5-72B-Instruct achieving comparable results. Notably, both models excelled in composition accuracy (0.89-0.90) and maintained consistent performance across Precision, Recall, and F1-score metrics (0.79-0.85). Llama-3.3-70B-Instruct showed strong overall performance with an accuracy of 0.76 and exceptional composition accuracy (0.87). Google's Gemini models presented mixed results. While Gemini-2.0-Flash achieved moderate performance with balanced metrics, Gemini-2.5-Flash-Preview unexpectedly underperformed compared to its predecessor, showing lower scores across most metrics (0.61-0.71). Llama-4-Maverick-17B-Instruct demonstrated notable strengths in specific areas despite its overall lower performance, achieving commendable composition accuracy (0.83) and Precision (0.78). However, the model struggled significantly with synthesis accuracy (0.55) and normalised Precision (0.63). The most concerning performance was observed with GPT-4.1-Nano, which consistently scored lowest across all metrics, particularly struggling with normalised Precision (0.46). Similarly, Gemma-3-27B-Instruct showed suboptimal performance, with notable weaknesses in synthesis accuracy and normalised Precision. 

\begin{figure}[pt!]
    \centering
    \includegraphics[width=0.78\linewidth]{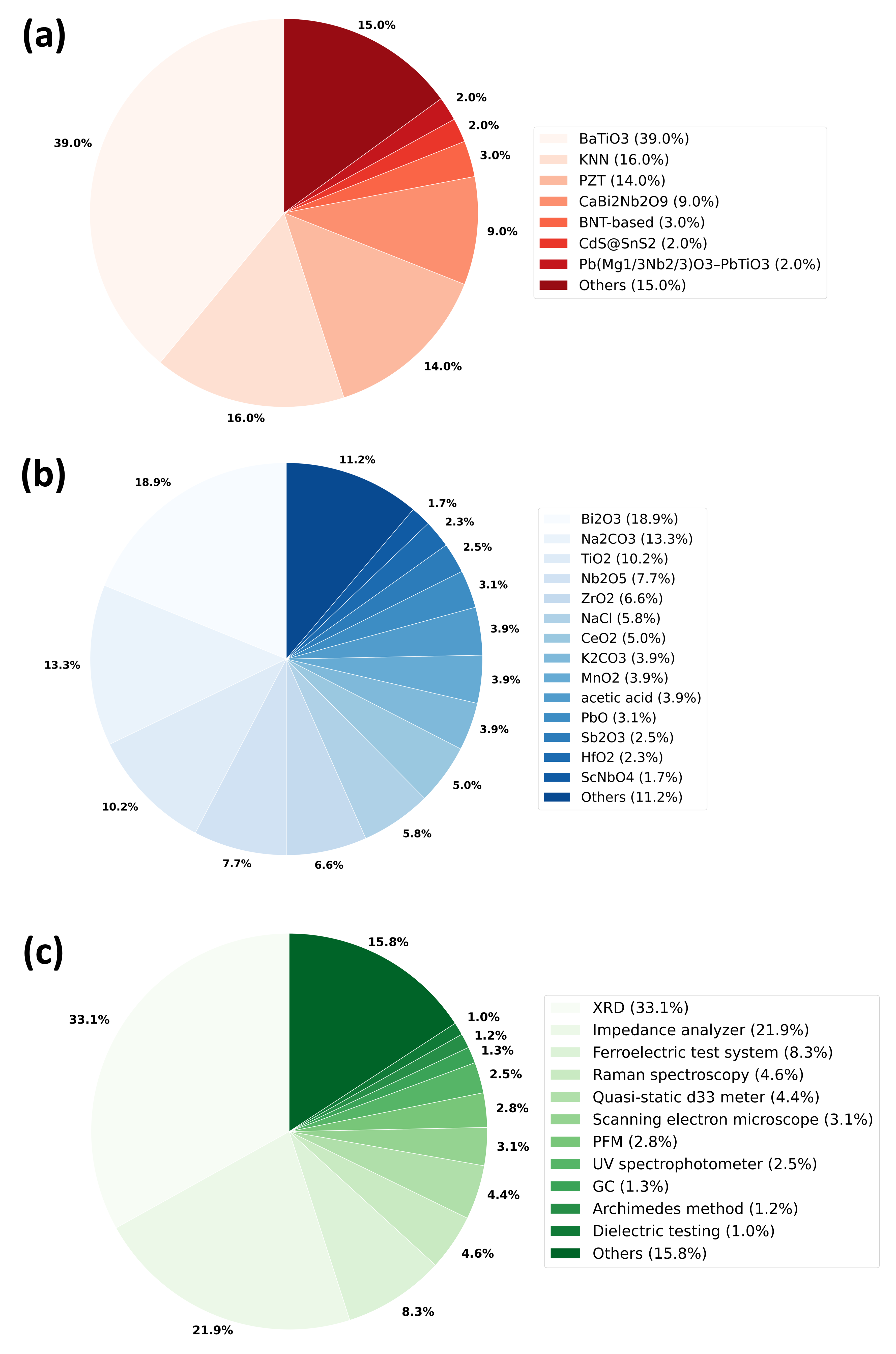}
    \caption{Distribution of various data types across the evaluated 100 articles: (a) piezoelectric material families, (b) synthesis precursors, and (c) characterisation techniques. Similarity thresholds of 0.8 were applied for families and precursors, whilst 0.78 was used for characterisation techniques to group semantically similar items.}
    \label{fig:distributions}
\end{figure}

Furthermore, to compare \textsf{ComProScanner}'s variable parsing ability with the original material-parsers tool developed by Foppiano \textit{et al.}\cite{foppiano2023automatic}, we tested several examples from the test dataset by processing them directly through material-parsers, with results summarised in \autoref{table:material-parsers-comparison}. Whilst \textsf{ComProScanner} outperformed material-parsers in most cases (first three examples), both tools successfully resolved the chemical formulae for relatively straightforward compositions (fourth example), and both occasionally failed, as demonstrated in the fifth example. Throughout the entire test set, \textsf{ComProScanner} demonstrated superior performance in most instances and equivalent performance in others when compared to material-parsers, thereby validating \textsf{ComProScanner}'s capabilities.

\begingroup
\renewcommand{\arraystretch}{1}
\renewcommand{\cellgape}{\Gape[2pt][2pt]}
\footnotesize
\begin{longtable}{|>{\centering\arraybackslash}m{2.85cm}|>{\centering\arraybackslash}m{2.25cm}|>{\raggedright\arraybackslash}p{10cm}|}
\caption{Comparison of performance between material-parsers developed by Foppiano \textit{et al.}\cite{foppiano2023automatic} and \textsf{ComProScanner} regarding variable substitution in material compositions.}
\label{table:material-parsers-comparison} \\
\hline
\textbf{DOI} & \textbf{Item} & \multicolumn{1}{>{\centering\arraybackslash}m{10cm}|}{\textbf{Details}} \\
\hline
\endfirsthead

\multicolumn{3}{c}%
{{\bfseries}} \\
\hline
\textbf{DOI} & \textbf{Item} & \multicolumn{1}{>{\centering\arraybackslash}m{10cm}|}{\textbf{Details}} \\
\hline
\endhead

\hline
\endfoot

\hline
\endlastfoot

\multirow{3}{*}{\makecell{10.1016/j.jallcom.\\2024.176609}} & Text & The 0.12Pb(Ni1/3Ta2/3)O3-xPbZrO3-(0.88-x)PbTiO3 piezoelectric ceramics with 2 mol\% MnO2 (abbreviated as PNT-xPZ-PT-Mn, x = 0.41, 0.42, 0.43, 0.44) were fabricated by the conventional solid-state reaction method \\
\cline{2-3}
& material-parsers & \makecell[tl]{\\[-25pt]1. 0.12Pb \\
2. Ni1/3Ta2/3)O2.59Pb\-ZrO3-(0.87.59)Pb\-TiO3\\
3. Ni1/3Ta2/3)O2.58Pb\-ZrO3-(0.87.58)Pb\-TiO3\\
4. Ni1/3Ta2/3)O2.57Pb\-ZrO3-(0.87.57)Pb\-TiO3\\
5. Ni1/3Ta2/3)O2.56Pb\-ZrO3-(0.87.56)Pb\-TiO3} \\
\cline{2-3}
& \textsf{ComProScanner} & \makecell[tl]{1. 0.12Pb(Ni1/3Ta2/3)O3-0.41PbZrO3-0.47PbTiO3 + 2\% MnO2 \\
2. 0.12Pb(Ni1/3Ta2/3)O3-0.42PbZrO3-0.46PbTiO3 + 2\% MnO2\\
3. 0.12Pb(Ni1/3Ta2/3)O3-0.43PbZrO3-0.45PbTiO3 + 2\% MnO2\\
4. 0.12Pb(Ni1/3Ta2/3)O3-0.44PbZrO3-0.44PbTiO3 + 2\% MnO2}\\
\hline
\multirow{3}{*}{\makecell{10.1016/j.\\jeurceramsoc.\\2025.117193}} & Text & In this study, dense Pb(1-x)K2x[Nb0.96Ta0.04]2O6 (PKxNT, x = 0.05, 0.10, 0.15, 0.20) ceramics were prepared via the solid-state reaction method. \\
\cline{2-3}
& material-parsers & \makecell[tl]{\\[-25pt]1. In \\
2. Pb(0.95)K20.05[Nb0.96Ta0.04]2O6\\
3. Pb(0.9)K20.10[Nb0.96Ta0.04]2O6\\
4. Pb(0.85)K20.15[Nb0.96Ta0.04]2O6\\
5. Pb(0.8)K20.20[Nb0.96Ta0.04]2O6} \\
\cline{2-3}
& \textsf{ComProScanner} & \makecell[tl]{1. Pb0.95K0.1[Nb0.96Ta0.04]2O6\\
2. Pb0.9K0.2[Nb0.96Ta0.04]2O6\\
3. Pb0.85K0.3[Nb0.96Ta0.04]2O6\\
4. Pb0.8K0.4[Nb0.96Ta0.04]2O6}\\
\hline
\multirow{3}{*}{\makecell{10.1016/j.ceramint.\\2024.09.282}} & Text & BaCO3 (99.8 \%, Aladdin), TiO2 (99.0 \%, McLean, Shanghai, China), SnO2 (99.9 \%, Aladdin), CaCO3 (99.0 \%, Sinopharm), Bi2O3 (99.9 \%, McLean), Fe2O3 (99.0 \%, Sinopharm) are used as raw materials, which were accurately weighed according to a composition of (1-x) (Ba0.95Ca0.05) (Ti0.89Sn0.11)O3-xBiFeO3 (BCTSO-xBFO, x = 0, 0.1, 0.5, 0.9 mol\%) and milled with ethanol for 16 h. \\
\cline{2-3}
& material-parsers & \makecell[tl]{\\[-25pt]1. BaCO3 \\
2. TiO2\\
3. SnO2 (99.9 \%, Aladdin\\
4. CaCO3 (99.0 \%, Sinopharm\\
5. Bi2O3 (99.9 \%, McLean), Fe2O3\\
6. (1.0) (Ba0.95Ca0.05) (Ti0.89Sn0.11)O3.0BiFeO3\\
7. (0.9) (Ba0.95Ca0.05) (Ti0.89Sn0.11)O2.9BiFeO3\\
8. (0.5) (Ba0.95Ca0.05) (Ti0.89Sn0.11)O2.5BiFeO3\\
9. (1.0) (Ba0.95Ca0.05) (Ti0.89Sn0.11)O3.0BiFeO3\\
10. (0.9) (Ba0.95Ca0.05) (Ti0.89Sn0.11)O2.9BiFeO3\\
11. (0.5) (Ba0.95Ca0.05) (Ti0.89Sn0.11)O2.5BiFeO3\\
12. (0.1) (Ba0.95Ca0.05) (Ti0.89Sn0.11)O2.1BiFeO3\\
13. (-15.0) (Ba0.95Ca0.05) (Ti0.89Sn0.11)O-13.0BiFeO3} \\
\cline{2-3}
& \textsf{ComProScanner} & \makecell[tl]{1. (Ba0.95Ca0.05)(Ti0.89Sn0.11)O3\\
2. (Ba0.95Ca0.05)(Ti0.89Sn0.11)O3 - (0.1)BiFeO3\\
3. (Ba0.95Ca0.05)(Ti0.89Sn0.11)O3 - (0.5)BiFeO3\\
4. (Ba0.95Ca0.05)(Ti0.89Sn0.11)O3 - (0.9)BiFeO3}\\
\hline
\multirow{3}{*}{\makecell{10.1016/j.ceramint.\\2024.10.314}} & Text & Lead-free piezoelectric ceramics with the formula Ba1-xSrxTi0.92Zr0.08O3 [x = 0, 0.04, 0.08, 0.12, 0.16, 0.2 (mol)] were prepared using the solid-state reaction technique. \\
\cline{2-3}
& material-parsers \& & \multirow{2}{*}[2ex]{\makecell[tl]{1. Ba1.0Sr0Ti0.92Zr0.08O3 \\
2. Ba0.96Sr0.04Ti0.92Zr0.08O3}}\\[20pt]
\cline{2-2}
& \textsf{ComProScanner} (resolved by both) & {\makecell[tl]{\\[-40pt]3. Ba0.92Sr0.08Ti0.92Zr0.08O3\\
4. Ba0.88Sr0.12Ti0.92Zr0.08O3\\
5. Ba0.84Sr0.16Ti0.92Zr0.08O3\\
6. Ba0.8Sr0.2Ti0.92Zr0.08O3}} \\
\hline
\multirow{4}{*}
{\makecell{10.1016/j.\\jeurceramsoc.\\2024.117065}} & Text & Pure CaBi2Nb2O9 and rare-earth thulium-substituted CaBi2Nb2O9 powders with nominal compositions of Ca1-xTmxBi2Nb2O9 (CBN-100xTm) were prepared through a solid-phase reaction method. To characterize the phase transition in detail, a composition range of x = 0.01–0.05 was selected.\\
\cline{2-3}
& material-parsers & \makecell[tl]{\\[-25pt]1. CaBi2Nb2O9\\
2. CaBi2Nb2O9\\
3. Ca1-xTmxBi2Nb2O9}\\
\cline{2-3}
& \textsf{ComProScanner} & \makecell[tl]{1. CaBi2Nb2O9 - 1Tm\\
2. CaBi2Nb2O9 - 2Tm\\
3. CaBi2Nb2O9 - 3Tm\\
4. CaBi2Nb2O9 - 4Tm\\
5. CaBi2Nb2O9 - 5Tm}\\
\cline{2-3}
& Actual resolved compositions & \makecell[tl]{\\[-40pt]1. Ca0.99Tm0.01Bi2Nb2O9\\
2. Ca0.98Tm0.02Bi2Nb2O9\\
3. Ca0.97Tm0.03Bi2Nb2O9\\
4. Ca0.96Tm0.04Bi2Nb2O9\\
5. Ca0.95Tm0.05Bi2Nb2O9}\\
\hline
\end{longtable}
\endgroup

\textsf{ComProScanner} also offers built-in data distribution visualisation functions to represent various material families, synthesis precursors, and characterisation techniques as either histograms or pie-charts through a semantic clustering mechanism. \autoref{fig:distributions} shows these data distributions, where similarity thresholds of 0.8 (default in \textsf{ComProScanner}) were applied for material families and precursors, while 0.78 was found to be best for characterisation techniques during semantic clustering. The resulting distributions reveal the prevalence of different components in piezoelectric materials research across the evaluated 100 articles. In terms of material families, \ce{BaTiO3} dominates at 39.0\%, followed by KNN (16.0\%) and PZT (14.0\%), with various other compositions including CaBi2Nb2O9 (9.0\%) and BNT-based materials (3.0\%) comprising the remainder. For synthesis precursors, Bi2O3 is most frequently used (18.9\%), followed by Na2CO3 (13.3\%) and TiO2 (10.2\%), with a diverse range of other precursors including various carbonates, oxides, and acids distributed across smaller percentages. The characterisation techniques show XRD as the predominant method (33.1\%), which is expected for crystalline phase analysis, followed by impedance analysers (21.9\%) for electrical property measurements, and ferroelectric test systems (8.3\%) for specific piezoelectric characterisation, with various other analytical techniques contributing to comprehensive materials evaluation. 

\begin{figure}[h]
    \centering
    \includegraphics[width=1\linewidth]{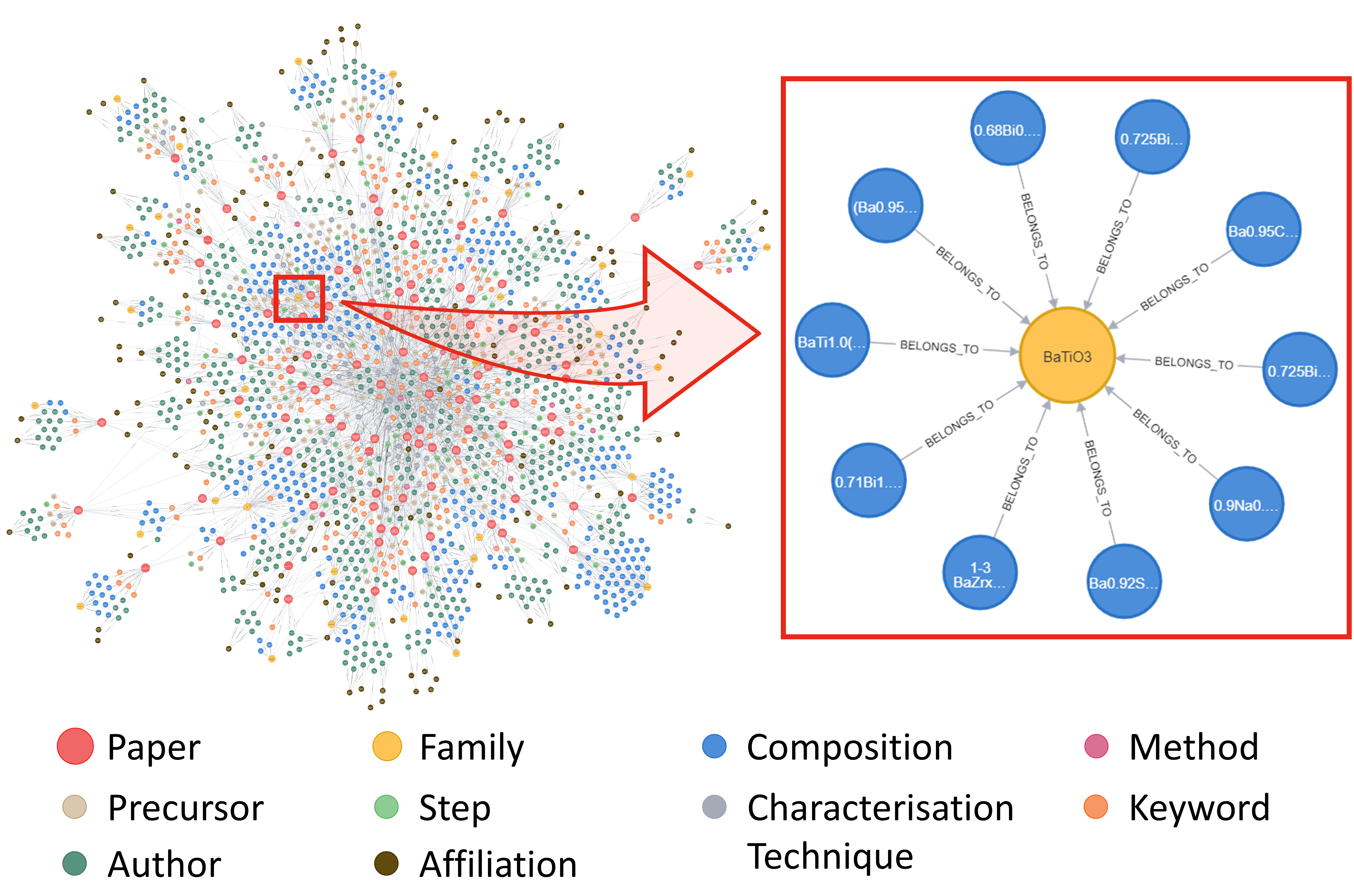}
    \caption{Generated \textit{neo4j} knowledge graph from both composition-property, and synthesis data as well as metadata information extracted from 100 articles using the \textsf{ComProScanner} package. The inset shows 10 randomly chosen items from the 79 compositions associated with \ce{BaTiO3} family node using cypher query.}
    \label{fig:knowledge_graph}
\end{figure}

To visualise the relationships between the distribution of all data types mentioned in the Methods section for the evaluated dataset, knowledge graph visualisation offered by the \textit{neo4j}\cite{neo4j2025} library has been employed within the \textsf{ComProScanner} package (\autoref{fig:knowledge_graph}). The produced \textit{neo4j} knowledge graph from 100 test articles contains a total of 1,825 graph nodes, which are summarised in Table S2 of the ESI. Cypher\cite{neo4j2025cypher} queries can be utilised to retrieve relational information for specific nodes, for example, the inset of \autoref{fig:knowledge_graph} represents 10 random items among 79 compositions associated with \ce{BaTiO3} family across 100 test articles. Detailed information about all nodes associated with the test data can be found in Table S2 in the ESI.

% Cypher\cite{neo4j2025cypher} queries can be utilised to retrieve relational information for specific nodes, for example, the
% \begin{minted}{cypher}
% MATCH (c:Composition)-[:BELONGS_TO]->(f:Family {name: "BaTiO3"}) 
% WITH c, f, rand() AS r 
% ORDER BY r 
% RETURN c, f LIMIT 10
% \end{minted}
% query retrieves 10 random items among 79 compositions associated with \ce{BaTiO3} family across 100 test articles which is illustrated in the inset of \autoref{fig:knowledge_graph}.

\subsection{Discussion}
Although \textit{d$_{33}$} was mentioned in 3,916 papers published within the considered time period, only data from the 100 test papers were extracted for evaluation, as our aim here is to introduce the robust data-extraction framework rather than providing a dataset. However, with this limited dataset of 100 test samples, we identified a piezoelectric composition Pb(In1/2Nb1/2)O3-Pb(Mg1/3Nb2/3)O3-PbTiO3 achieving 2090 pC/N, which demonstrates the significance of this framework. On top of this, over 99\% of the extracted piezoelectric materials are not in the Materials Project piezoelectric database, which emphasises the importance of \textsf{ComProScanner} for creating datasets from materials information buried within the materials science literature. When using \textsf{ComProScanner} to extract data for a use case different to the piezoelectric materials described here, substituting the piezoelectric material and \textit{d$_{33}$} coefficient-related keywords with one's own choice of property keywords in the additional information maybe sufficient; however, one may need to introduce further prompt engineering for better extraction performance. During evaluation with one's own data, despite being superior in evaluation, the agentic approach can result in substantial costs, as the pricing of reasoning models is considerably higher than that of chat models. In such cases, to determine the suitable model for data extraction, semantic evaluation can serve as a cost-efficient approach. 

For the piezoelectric materials considered, balanced performance in each metric with an overall accuracy of 0.82 indicates that DeepSeek-v3-0324 possesses the most reliable extraction capabilities for complex piezoelectric material data. The consistency in various metrics for both Qwen models suggests these models are also well-suited for systematic materials data extraction tasks. Llama-3.3-70B-Instruct's results makes it particularly valuable for applications requiring high Precision in materials identification. However, its synthesis accuracy (0.65) is relatively lower, indicating potential challenges in extracting complex synthesis information. The counter-intuitive results from two Gemini models suggests that model updates do not always guarantee improved performance for domain-specific tasks. The results for Llama-4-Maverick-17B-Instruct suggest it may be more suitable for composition-focused extraction tasks rather than comprehensive materials informatics applications. Poor performances from GPT-4.1-Nano and Gemma-3-27B-Instruct highlight the importance of model selection for materials informatics applications, where domain-specific performance can vary significantly from general language tasks. The comparison between the original material-parsers tool and ComProScanner demonstrates that our package performs significantly more efficiently than material-parsers when resolving complex chemical compositions containing variables.

Though LLM agents attempt to ensure consistent results across runs, the underlying LLMs are nondeterministic by nature, which forms the core limitation of any type of LLM-based approach; consequently, results may vary slightly between runs. For the \textit{Materials Data identifier} agent, the RAG question, chunk size, chunk overlap, top k value, and RAG chat model may require adjustment and testing according to the specific use case. Although manual evaluation would serve as the optimal evaluation technique compared to semantic and agentic approaches, it is not practical for large dataset evaluation. With this consideration, semantic and agentic approaches are incorporated into the framework, and depending on the chosen reasoning model, evaluation results can vary slightly. 

 % Currently, as no Optical Character Recognition (OCR) technology or vision-language model (VLM) is integrated with the framework, it cannot extract data presented in graphs or other image formats. Finally, whilst some data cleaning tools are included in the package, users may need to implement their own pre-processing steps to construct a final dataset suitable for machine-learning model development. Presently, the extracted data format is configured to accommodate only one materials property extraction structure. If users wish to extract multiple properties associated with a single material composition, they must modify the package to incorporate the expected JSON structure.

\textsf{ComProScanner} establishes the essential foundation for the next generation of AI in material science, creating a pathway to develop extensive text-mined datasets from journal articles. The framework we have developed enables a seamless, user-friendly automated data extraction pipeline. However, OCR technology or VLMs could be integrated with the framework in the future to extract information from graphs or other image formats. Additionally, flexibility to modify the structure of the extracted JSON data could be incorporated into the framework to extract multiple material properties.

\subsection{Conclusions}
Although researchers have attempted to automate the extraction of structured information from journal articles, a user-friendly, ready-to-use framework was lacking. Here, we have introduced a multi-agent framework, \textsf{ComProScanner} to accomplish this task. We assessed our framework using 100 scientific articles across 10 LLMs for highly complex ceramic piezoelectric material compositions and corresponding \textit{d$_{33}$} coefficient values. We found \textsf{ComProScanner} could extract extremely complex piezoelectric material compositions with the optimal settings with DeepSeek-V3-0324 achieving an overall accuracy of 0.82 and compositional accuracy of 0.90. Both considered Qwen models (Qwen3-235B-A22B and Qwen-2.5-72B-Instruct) and Llama-3.3-70B-Instruct also demonstrated competitive performance with an average composition-property accuracy ranging from 0.87-0.90. Surprisingly Gemini-2.5-Flash-Preview underperformed compared to its predecessor in most of the evaluation metrics. Correct assessment of these complex textual data is challenging, and must be performed manually which is practically impossible for extensive datasets. However, both semantic and agentic evaluation suggest the potential application of \textsf{ComProScanner} for creating vast datasets of complex material compositions and associated properties, along with synthesis information. As demonstrated by \textsf{ComProScanner}'s performance, LLM-based multi-agent frameworks represent a promising approach for automated scientific data extraction, potentially accelerating materials discovery and database construction for data-driven research.

\subsection{Code and Data availability}

\textsf{ComProScanner} code is available at \url{https://github.com/slimeslab/ComProScanner} for reuse and modification under the \href{https://github.com/slimeslab/ComProScanner/blob/main/LICENSE}{MIT licence}. The Python package is hosted on the Python Package Index (PyPI) at \url{https://pypi.org/project/comproscanner/} for straightforward installation via pip. Comprehensive documentation detailing package usage with custom configurations and all available functions with their accepted arguments is available at \url{https://slimeslab.github.io/ComProScanner}. All data pertaining to the evaluation process can be found in the \href{https://github.com/slimeslab/ComProScanner/tree/main/examples}{examples} folder.

\subsection{Author Contributions}

Aritra Roy: conceptualisation, data curation, formal analysis, investigation, methodology, software, validation, writing – original draft. Enrico Grisan: resources, supervision. John Buckeridge: conceptualisation, formal analysis, funding acquisition, investigation, resources, validation, writing – original draft, writing – review \& editing, supervision. Chiara Gattinoni: conceptualisation, formal analysis, funding acquisition, investigation, resources, validation, writing – original draft, writing – review \& editing, supervision.

\subsection{Conflicts of Interest}

There are no conflicts to declare.

\begin{acknowledgement}

A.R. and J.B. thank London South Bank University for financial and legal support to obtain Elsevier, Wiley, and Springer Nature publisher's TDM licences. C.G. thanks King's College London for legal support in obtaining IOP Publishing's TDM licence and for funding the Article Processing Charge (APC) to publish the journal as an open-access article.

\end{acknowledgement}
    
% \begin{suppinfo}

% This will usually read something like: ``Experimental procedures and
% characterization data for all new compounds. The class will
% automatically add a sentence pointing to the information on-line:

% \end{suppinfo}

\bibliography{references}

\end{document}